\documentclass[aps,twocolumn,prd,nofootinbib]{revtex4}
\usepackage{amsmath}
\usepackage{graphicx}
\usepackage{dcolumn}
\usepackage{bm}
\usepackage{amssymb}
\usepackage{latexsym}

\begin{document}

\title{Orbital Motion and Quasi-quantized Disk around Rotating Neutron Stars}

\author{Joan Jing Wang\footnote{Email address: jwang@mx.nthu.edu.tw}, Hsiang-Kuang Chang}
\affiliation{Institute of Astronomy, National Tsing Hua University,
Hsinchu 30013, Taiwan}
\affiliation{Department of Physics, National Tsing Hua University, Hsinchu
30013, Taiwan}

\begin{abstract}
In accreting neutron star (NS) low-mass X-ray binary (LMXB) systems, 
NS accretes material from its low-mass companion via a Keplerian disk. 
In a viscous accretion disk, inflows orbit the NS and spiral in 
due to dissipative processes, such as the viscous process and collisions of elements. 
The dynamics of accretion flows in the inner region of an accretion disk 
is significantly affected by the rotation of NS. 
The rotation makes NS, thus the space-time metric, 
deviate from the originally spherical symmetry, and leads to gravitational quadrupole, 
on one hand. On the other hand, a rotating NS drags 
the local inertial frame in its vicinity, 
which is known as the rotational frame-dragging effect. 
In this paper, we investigate the orbital motion of accretion flows of accreting NS/LMXBs and demonstrate that the rotational effects of NS 
result in a band of quasi-quantized structure 
in the inner region of the accretion disk, which is different, in nature, from the scenario in the strong gravity of black hole arising from the resonance for frequencies related to epicyclic and orbital motions. We also demonstrate that such a disk structure may account for 
frequencies seen in X-ray variability, such as quasi-periodic oscillations, and can be 
a potential promising tool of investigation for photon polarization.
\end{abstract}

\maketitle

\section{Introduction}

%General description of accretion scenario in NS/LMXBs
In neutron star (NS) low mass X-ray binaries (LMXBs), 
the low-mass companion 
is in a  phase called the Roche lobe overflow, 
during which material passes from the Roche lobe of the companion star 
to the NS through the inner Lagrange point \cite{2002apa..book.....F}. 
Within the Roche lobe of NS, the dynamics of transferred material is controlled 
dominantly by the gravitational field of the NS. 
In the spherically symmetric gravitational field of NS, 
a circular orbit has the least energy for a given angular momentum. 
Therefore, the dissipation in the orbiting gas tends to circularize its motion. 
If not too much angular momentum is lost during this process, 
the captured matter will orbit the NS in the binary plane and move in
 some circular orbits, at a Keplerian angular velocity 
$\omega_K(r) = \sqrt{\frac{GM}{r^3}}$ \cite{1973ApJ...184..271L}, 
where $G$ is the gravitational constant, $M$ represents the mass of the NS, 
and $r$ is the orbital radius of orbiting matter. 
With more and more material captured by the gravitational field of the NS, 
the accretion disk becomes compact enough, in which turbulence and clumps appear. 
In the turbulent and viscous Keplerian disk, 
the inflows should transport the angular momentum outward effectively 
and spiral in consecutively, in order to be accreted by NS \cite{1972A&A....21....1P}. 
This process resorts to the dissipative process, 
e.g. collisions of gas elements, shocks, viscous dissipation, and so on, 
which converts part of kinetic energy into internal energy, 
radiated as X-rays \cite{1978ApJ...223L..83G}.

%Stable orbits in vicious disk 
Because of the strong magnetic field, the accretion disk cannot 
extend directly to the surface of NS. Instead, the disk stops at a special radius \cite{1977ApJ...215..897E}. 
This is the innermost boundary of the accretion disk, corresponding to a "ledge-like" extremum of the effective potential.
During accretion process, inflows will experience some sub-effective potentials, which consist of the total potential of NS, in a turbulent and inhomogeneous accretion disk.  
The accreted matter, resting in the minimum of each effective potential, moves in stable circular motion at some particular radii \cite{Rothman:2000bz} due to inhomogeneities and turbulence, and oscillates around each radius, if slightly disturbed. 

%Holonomy in curved space-time 
The orbital motion of accretion flows controlled by 
the total potential of the gravitational and magnetic fields of the rotating NS 
is different from that in a flat space-time. 
As we know, the parallel transport of vectors along closed curves in a flat space-time 
can return the original vector, which is known as the holonomy invariance. 
However, in a curved space-time, a vector under parallel transport along a closed curve, 
in general, results in a deficit angle between the directions 
of the initial and final vectors after one loop. 
That is to say, the physical orbital motion of accretion flows 
cannot follow closed circular orbits. 
However, the holonomy invariance of physical orbits in a curved space-time 
can be satisfied under certain conditions \cite{Rothman:2000bz}. 
As an example, for the geodesic motion of a test particle along a circle with 
a constant radius in Schwarzschild geometry, the holonomy invariance occurs 
at some particular radii, after an integer number of loops \cite{Abramowicz:2002di}. 
The orbital motions of accreted matter on these radii correspond to the stable circular motions \cite{1998bhad.conf.....K, 1999PhR...311..259W},   
which can lead to quantization of orbits.
In addition, non-geodesic charged particles also can move along ultra-relativistic circular trajectories in the Kerr space-time of rotating black hole \cite{1981GReGr..13..899A} due to the strong gravitational field. The resonance between relativistic orbital Keplerian frequency and related epicyclic frequencies in accretion disk of orbiting Kerr black holes leads to some trapping effects, which can explain the observed high frequency quasi-periodic oscillations (QPOs) with given ratios \cite{2005A&A...437..775T}.

%Rotation effects of NS and Organization of the paper
{\bf However, the gravitational field of a canonical NS is weaker than that of black hole, because of the relatively low surface gravitational redshift of NS \cite{2007ASSL..326.....H}. As a result, the relativistic effects \cite{Rothman:2000bz, Abramowicz:2002di, 1981GReGr..13..899A, 2005A&A...437..775T} in accretion disk arising from strong gravity of both rotating and non-rotating black hole cannot play a corresponding significant role in the scenario of NS.} That is to say, the gravitational field of NS is difficult to lead to significantly trapping effects in the inner region of accretion disk described in that of black hole \cite{Abramowicz:2002di}.
The accreted material controlled by the gravitational field of NS follows classical motion. Besides, the rotation effects of NS on the motions of charged particles become important.
In this paper, we investigate the rotational effects of NS on accretion flows in weak field approximation, in order to interpret the formation mechanism for stable circular motion of accreted flow and the structure of the inner accretion disk. 
Our work resorts to Hamiltonian mechanics to describe the classical motion 
of accretion material in the weak gravitational field and strong magnetic field of NS. 
In addition, the rotation of NS gives rise to two effects. 
Firstly,  stellar oblateness arises from the rotation 
and thus a quadruple term in the gravitational potential appears \cite{1973grav.book.....M}. 
Secondly, a rotational massive object will impose a rotational frame-dragging effect 
on the local inertial frame, which is known as gravitomagnetism 
\cite{1918PhyZ...19...33T, 1918PhyZ...19..156L, 1921PhyZ...22...29T, 1984GReGr..16..711M}, 
according to the general theory of relativity. 
Considering that the gravitomagnetism is an effect of weak gravitational field because of a perturbation in space-time metric, and that 
NSs in LMXBs are non-relativistic rotators \cite{ATNF}. 
We, therefore, treat the oblateness and gravitomagnetism arising from the rotation of NS 
as a perturbation in the Hamiltonian and study their influence on the orbital motion of accretion flow.
%which is responsible for a quasi-quantized band of orbits in the inner region of the accretion disk. 
In section II, we present the Hamiltonian mechanics of test particles 
in a system consisting of accreted plasma and controlled 
by gravitational and strong magnetic fields of NS in LMXBs. 
We review the gravitoelectromagnetism (GEM) and study the behavior of accretion flows in both gravitoelectromagnetic field and axisymmetric space-time in Section III. 
The closure and quasi-quantization condition and closure circular orbits 
are demonstrated in section IV. 
Section V contains summary and discussion.

\section{Hamiltonian and Classical motion}

We consider a system consisting of accreting plasma, 
which contains charged particles, i.e. positive ions, electrons and negative ions, 
in the framework of 4-dimension space-time. For a test particle, 
there are 4 generalized coordinates $q_1$, $q_2$, $q_3$, and $q_0$, 
i.e. with 4 degrees of freedom. 
The first three describe the space coordinates $x$, $y$, $z$ (or $r$, $\theta$, $\phi$), 
and $q_0$ denotes the time component $t$. 
Accordingly, the generalized coordinates and generalized velocities 
can be written as $q_i$ and $\dot{q}_i = dq_i/dq_0$ 
for the $i$th degree of freedom\footnote{Note that we choose Greek subscripts 
and indices (i.e. $\mu$, $\nu$, $\alpha$, $\beta$) 
to describe the 4-dimension space-time components (0, 1, 2, 3) for test particles,
 while  the subscripts and indices $i$, $j$, $k$ denote 
the 3-dimension space coordinates $x$, $y$, $z$ (or $r$, $\theta$, $\phi$).}, respectively. 
We introduce canonically conjugate variables to the $q_i$, 
which are called the generalized momenta $p_i$. 
Therefore, the mechanics of the system can be described in terms of 
the Hamiltonian $H$ \cite{1965Mechan.book},
which depends on 
the canonical coordinates $q_i$, generalized momenta $p_i$ and time $q_0$: 
\begin{equation}
H(q_i, p_i, q_0) = {\sum\limits_{i=1}^{f} p_i \dot{q}_i} - L(q_i, \dot{q}_i, q_0),
\end{equation}
where $L(q_i, \dot{q}_i, q_0)$ is the Lagrange function of the system 
and $p_i = \partial L/\partial \dot{q_i}$ \cite{1950classi.book}. 
Consequently, the equations of motion, i.e. Hamiltonian equations of motion, 
can be obtained from the Lagrange equations of motion 
$\frac{d}{dq_0}(\frac{\partial L}{\partial\dot{q_i}})-\frac{\partial L}{\partial q_i} = 0$ 
\cite{1965Mechan.book} of the system, by the aid of Legendre transformations,
\begin{eqnarray}
\dot{q}_i &=& \frac{\partial H}{\partial p_i},\\
\dot{p}_i &=& -\frac{\partial H}{\partial q_i}
\end{eqnarray}
The Hamiltonian $H = T + V$ denotes the total energy of the system, 
with kinetic energy $T$ and potential $V$.

In a plasma disk, the charged particles in the magnetic and gravitational fields of 
a NS are subject to the Lorentz force and gravity, 
so we have\footnote{In this work, we set the light speed c=1.}
\begin{equation}
%m \ddot{\vec{r}} = q\frac{\dot{\vec{r}}}{c} \times \vec{B} - \frac{GMm}{r^3}\vec{r},
m \ddot{q_i} = Q \dot q_j B_k \varepsilon^{jk}_{i} - \frac{GMm}{q^2_i},\label{force}
\end{equation}
where $m$ and $Q$ are the mass and charge of test particle, 
$B_k$ is the $k$th component of the magnetic field, 
and $\varepsilon_{ijk}$ is the 3-dimentional completely antisymmetric tensor 
of Levi-Civita. Defining the canonical momenta of charged particles 
as $\pi_i = p_i - Q A_i$, where $p_i$ is the generalized physical momenta, 
and $A_i$ is the vector potential of the magnetic field 
$B_i = \bigtriangledown_j A_k \varepsilon^{jk}_{i}$, 
the Hamiltonian reads 
\begin{equation}
H = \frac{1}{2m}\sum\limits_i[(p_i - QA_i)^2 + \frac{GMm}{q_i}].\label{Hamil}
\end{equation}
As a result, the equations of motion for charged particles, 
without perturbation, in the gravitational and magnetic fields of a NS are written as
\begin{eqnarray}
\dot{q}_i &=& \frac{p_i-QA_i}{m},\label{cor-Ham}\\
\dot{p}_i &=& -\frac{Q}{m}\sum_j[{p_j-QA_j}] \frac{\partial A_j}{\partial q_j}\delta_{ij}+ \frac{GMm}{q^2_i}.\label{mom-Ham}
\end{eqnarray}
The motion of accretion flows following 
Eq. (\ref{cor-Ham}) and (\ref{mom-Ham}) corresponds 
to a helicoid, with uniformly open circular motion at different radii.

\section{Rotational effects}

NSs are rotating objects, and the rotation of NS has influence on the behavior of accretion flows
 in its vicinity. In analogy with electromagnetism described by Maxwell's equations, 
the rotational gravitational mass can give rise to GEM. 
Moreover, the rotation of NS will break the spherical symmetry 
and produce axisymmetric space-time. In this section, 
we investigate the behavior of accretion flows in the gravitoelectromagnetic field 
and axisymmetric space-time in the vicinity of a rotating NS. 

\subsection{Gravitoelectromagnetism}

GEM is based on the close formal analogy between Newton's law of gravitation 
and Coulomb's law of electricity. 
The Newtonian solution of the gravitational field can be alternatively 
interpreted as a gravitoelectric field. It is well known that a magnetic field 
is produced by the motion of electric charges, i.e. electric current. 
Accordingly, the rotating mass-current would give rise to a gravitomagnetic field. 
In the framework of the general theory of relativity, 
a non-Newtonian massive mass-charge current can produce a gravitoelectromagnetic field 
and lead to interesting physical properties 
\cite{1950relati.book, 2001referGM.book....BJ, 2001referGM.book....M}.

Due to the relatively low surface gravity of NS \cite{2007ASSL..326.....H}, 
we can use the linear approximation of the gravitational field, 
with non-relativistic rotation. A global background $g_{\mu\nu}$ 
of the inertial frame with coordinates $q_{\mu} = (q_0, q_i)$ 
and Minkowski metric $\eta_{\mu\nu}$ is perturbed due to the presence 
of gravitating source with a perturbative term $h_{\mu\nu}$, according to
\begin{equation}
g_{\mu\nu} = \eta_{\mu\nu} + h_{\mu\nu},~~~|h_{\mu\nu}| \ll 1 \,\, .
\end{equation}
We define the trace-reversed amplitude 
$\bar{h}_{\mu\nu} = h_{\mu\nu} -\frac{1}{2}\eta_{\mu\nu}h$, 
where $h = \eta^{\mu\nu}h_{\mu\nu} = h^\alpha_\alpha$ is the trace of $h_{\mu\nu}$.
Then, expanding the Einstein's field equations 
$G_{\mu\nu} = 8 \pi G T_{\mu\nu}$ in powers of $\bar{h}_{\mu\nu}$ 
and keeping only the linear order terms, 
we obtain the field equations (\cite{1973grav.book.....M} sec. 18.1)
\begin{equation}
\Box \bar{h}_{\mu\nu} = - 16\pi G T_{\mu\nu},\label{LEFE}
\end{equation}
where the Lorentz gauge condition $\bar{h}^{\mu\nu}_{~~,\nu} = 0$ is imposed. 
In analogy with the Maxwell's field equations $\Box A^{\nu} = 4 \pi j^{\nu}$, 
we can find that the role of the electromagnetic vector potential $A^{\nu}$ 
is played by the tensor potential $\bar{h}_{\mu\nu}$, 
while the role of the 4-current $j^{\nu}$ is played 
by the stress-energy tensor $T_{\mu\nu}$. 
Therefore, we can write down the solution of Eq. (\ref{LEFE}) 
in terms of a retarded potential \cite{2001LNP...562...83M, Ruggiero:2002hz}
\begin{equation}
\bar{h}_{\mu\nu} = 4G\int\frac{T_{\mu\nu}(q_0-|q_i-q'_i|,q'_i)}{|q_i-q'_i|}d^3q'_i.
\end{equation}

We are interested in the tensor potential $\bar{h}_{00}$ and $\bar{h}_{0i}$, 
and negelect all terms of high order and smaller terms including 
the tensor potential $\bar{h}_{ij}(q_0,q_i)$ \cite{1973grav.book.....M}. 
The explicit expression for the tensor potential $\bar{h}_{\mu\nu}$ is \cite{Ruggiero:2002hz}
\begin{eqnarray}
\bar{h}_{00} &=& 4\Phi,\\
\bar{h}_{0i} &=& -2A_i.\label{potential}
\end{eqnarray}
Here, $\Phi$ and $A_i$ are the Newtonian or gravitoelectric potential 
and the gravitomagnetic vector potential, respectively. They can be expressed as 
\begin{eqnarray}
\Phi(q_0,q_i) &=& -\frac{GM}{r},\label{GEP}\\
A_i(q_0, q_i) &=& G\frac{J_jq_k}{r^3}\varepsilon^{jk}_i,\label{GMP}
\end{eqnarray}
where $A_i$ is in terms of the angular momentum $J_i$ of NS. 
Accordingly, the Lorentz gauge condition reduces to
\begin{equation}
\frac{\partial\Phi}{\partial q_0} + \frac{1}{2}\bigtriangledown_iA_j\delta_{ij} = 0.\label{Lorgau}
\end{equation}

Then we can define the gravitoelectromagnetic field, 
i.e. the gravitoelectric field $g_i$ and the gravitomagnetic field $\Re_i$, 
as \cite{2001LNP...562...83M}
\begin{eqnarray}
g_i &=& - \bigtriangledown_i\Phi - \frac{\partial}{\partial q_0}(\frac{1}{2}A_j)\delta_{ij},\label{GEfield}\\
\Re_i &=& \bigtriangledown_jA_k\varepsilon^{jk}_i.\label{GMfield}
\end{eqnarray}
In the approximation of a weak field \cite{1973grav.book.....M, 1972gcpa.book.....W} 
and slow rotation for NS, $g_i$ contains mostly first order corrections 
to flat space-time and denotes the Newtonian gravitational acceleration, 
whereas $\Re_i$ contains second order corrections and is related to the rotation of NS.   

Using Eqs. (\ref{LEFE}), (\ref{Lorgau}), (\ref{GEfield}), 
and (\ref{GMfield}) and analogy with Maxwell's equations, 
we get the gravitoelectromagnetic field equations \cite{2001LNP...562...83M}
\begin{eqnarray}
&&\bigtriangledown_ig_j\delta_{ij} = -4 \pi G \rho_g,~\bigtriangledown_jg_k= -\frac{\partial}{\partial q_0}(\frac{1}{2}\Re_i),\nonumber\\
&&\bigtriangledown_i\Re_j\delta_{ij} = 0,~~~~~~\bigtriangledown_j\Re_k = \frac{\partial g_i}{\partial q_0} - 4\pi Gj_{gi}.
\end{eqnarray}
Here, $\rho_g$ is the mass density, and $j_g = \rho_g \times$ 
(velocity of the mass flow generating the gravitomagnetic field) 
denotes the mass current density or mass flux. 
These equations include the conservation law 
for mass current 
$\frac{\partial\rho}{\partial q_0} + \bigtriangledown_ij_{gj}\delta_{ij} = 0$, 
as they should be.

\subsection{Rotating relativistic stars and axisymmetric structure}

As described above, the effect of a rotating NS on space-time is referred to 
as the dragging of local inertial frame. On the other hand, 
the rotation of local inertial frames has a real effect on 
the inertial structure of the rotating star. 
The line element $ds^2 = g_{\mu\nu}dq^{\mu}dq^{\nu}$ for a rotating NS 
must possess some off-diagonal elements, which means that the NS is centrifugally deformed. 
We expect the rotation to centrifugally flatten the star more or less 
depending on its angular velocity. 
As a result, the spherical symmetry is broken, 
retaining only axial symmetry. In addition, we assume the NS to be static, 
considering that it is in uniform rotation and static in the co-rotating frame. 

In the general theory of relativity, the stationary and axisymmetric space-time metric, 
employed to describe the space-time geometry of a rotating star in equilibrium, 
has a form dictated by time-translational invariance and axial-rotational 
invariance\footnote{For convenience, we use polar coordinates $r$, $\theta$, $\phi$ 
and time $t$ to describe the space-time components in this part.} \cite{1971ApJ...167..359B},
\begin{eqnarray}
&&ds^2  = e^{2\nu(r,\theta)}dt^2 - e^{2\lambda(r,\theta)}dr^2\nonumber\\
&&- e^{2\mu(r,\theta)}[r^2d\theta^2 + r^2\sin^2\theta(d\phi - \omega(r,\theta)dt)^2].
\end{eqnarray}

In this metric, the non-diagonal elements appear as a new feature of axisymmetric structure,
\begin{equation}
g_{t\phi} = g_{\phi t} = r^2\sin^2\theta e^{2\mu(r,\theta)}\omega(r,\theta).
\end{equation}
Here, we are interested in the function $\omega(r)$. 
%The axisymmetric structure and the metric should be the same under  reversal of time as well as under  reversal of the angular velocity. So $\omega$ can depend only on odd powers of the angular velocity of NS. 
Considering a test particle at a great distance from NS in its equatorial plane of rotation ($\theta = \pi/2$, for convenience), we imagine that the particle drops from rest. Therefore, all three contravariant velocity components ($u^{r}$, $u^{\theta}$, $u^{\phi}$) of the particle are initially zero. In an axisymmetric space-time, $u_{\phi}$ is conserved and thus $u_{\phi} = 0$. Moreover, the $\phi$ coordinate of particle at $\theta = \pi/2$ obeys the equation \cite{1967ApJ...150.1005H},
\begin{equation}
\frac{d\phi}{dt} = \frac{g^{\phi t}}{g^{tt}} = \frac{-g_{\phi t}}{g_{\phi\phi}} = \frac{-r^2e^{2\mu(r,\theta)}\omega(r)}{-r^2e^{2\mu(r,\theta)}} = \omega(r).
\end{equation}
%would fall from rest at a great distance 
%from a non-rotating NS directly toward its center. 
%However, the particle will experience an increasing drag in the direction of rotation of NS. 
Therefore, $\omega(r)$ is referred to as the angular velocity of the local inertial frame, which 
%{\bf i.e. the angular velocity of accreted plasma}. 
can be expressed in terms of the Keplerian velocity at radius $r$ \cite{2003LRR.....6....3S},
\begin{equation}
\omega(r) = \frac{2}{5}\sqrt{\frac{GM}{r^3}}R^2\Omega^2,
\end{equation}
where $R$ denotes the radius of NS, and $\Omega$ is the angular velocity  of NS. Consequently, the accreted matter experiences an increasing drag in the direction of rotation of NS and possesses a corresponding angular velocity $\omega(r)$. 

\section{Closure condition and Quasi-quantization of the accretion disk}

Considering the gravitational field and magnetic field of NS, 
the mechanics of charged particles in the accretion disk is controlled 
by Eq. (\ref{Hamil}), which describes a helicoid motion. 
However, the GEM and axisymmetric metric, arising from the rotation of NS, imposes a perturbation on the Hamiltonian, which have influence on the helical orbital motion of accretion flows.

\subsection{Gravitational Larmor frequency}

The linear-order terms of Einstein's field equations in the form 
of Eq. (\ref{LEFE}) correspond to the solutions described 
by Eq. (\ref{GEP}) and (\ref{GMP}) around a rotating object. 
The corresponding space-time metric dominated by the gravitoelectromagnetic field 
arising from a rotating NS can be written as \cite{2001LNP...562...83M} 
\begin{eqnarray}
ds^2 &=& -(1 - 2\Phi)dq_0^2 - 4(A_idq_j\delta_{ij})dq_0 \nonumber \\  
&&+ (1+ 2\Phi)\delta_{ij} dq^i dq^j.\label{GEMmetric}
\end{eqnarray} 

The motion of a test particle $m$ in a gravitoelectromagnetic field follows 
from the Lagrangian $L = -m\frac{ds}{dq_0}$. 
According to Eq. (\ref{GEMmetric}), we obtain the equation of motion 
for test particles from the variational principle $\delta\int Ldq_0 = 0$,
\begin{equation}
L = -m(1-\dot{q}_i^2)^{\frac{1}{2}}+m\gamma(1+\dot{q}_i^2)\Phi-2m\gamma\dot{q}_iA_j\delta_{ij},\label{Lagrangian}
\end{equation}
where $\gamma$ is the Lorentz factor.

In fact, the geodesic motion of a particle, in the gravitoelectric field 
and dipolar gravitomagnetic field of NS with present formalism, 
can be cast in the form of an equation of motion under the action of a Lorentz force. 
The geodesic equation is
\begin{equation}
\frac{d^2q^{\alpha}}{ds^2} + \Gamma^{\alpha}_{\mu\beta}\frac{dq^{\mu}}{ds}\frac{dq^{\beta}}{ds} = 0,
\end{equation}
where $\Gamma^{\alpha}_{\mu\beta}$ are Christoffel symbols, i.e. affine connection.
For a test particle in non-relativistic motion, 
corresponding to a viscous accretion disk, 
we have $\frac{dq^0}{ds} \simeq 1$ and $\dot{q}_i \simeq \frac{dq^{\mu}}{ds}$, 
and we neglect the terms of higher orders for velocity. 
Limiting to static fields, where $g_{\mu\nu,0} = 0$, 
we can get the deviation of Eq. (\ref{Lagrangian}) 
from a free-particle Lagrangian given by the GEM force in GEM equation, in analogy with Lorentz force in electromagnetic field,
\begin{equation}
F_{GEMi} = Q_g g_i + Q_{\Re}\dot{q}_j\Re_k\varepsilon^{jk}_i,\label{Lorentz}
\end{equation}
where $Q_g = -m$ and $Q_{\Re} = -2m$ are the gravitoelectric charge 
and gravitomagnetic charge of the test particle \cite{2001LNP...562...83M}, 
respectively. %It follows from the geodesic motion of a test particle of mass $m$ far from the source in the rotating gravitational background that the canonical momentum of the particle is given approximately by $\pi_i = p_i + (-2mA_i)$, where $p_i$ is the kinetic momentum. Correspondingly, the attractive nature of gravity is reflected by negative gravitational charges for the test particle. The gravitomagnetic charge is always twice the gravitoelectric charge as a consequence of the tensorial character of the gravitational potentials in the general theory of relativity.

In analogy with electromagnetism, the Larmor quantities in 
the gravitoelectromagnetic field would be 
\begin{eqnarray}
\vec{a}_{L} &=& -Q_g\vec{g}/m,\label{traacc}\\
\vec{\omega}_{L} &=& Q_{\Re}\vec{\Re}/2m.\label{rotafre}
\end{eqnarray}
Therefore, the test particle has the translational acceleration 
$\vec{a}_{L} = \vec{g}$ and the rotational frequency $\omega_L = |\vec{\Re}|$.

For a rotating NS with angular velocity $\Omega$, the gravitomagnetic potential, thus the gravitational Larmor frequency of test particles 
in a gravitoelectromagnetic field, can be written as \cite{Mirza:2003fq}
\begin{equation}
\omega_L = |\vec{\Re}| = \frac{4}{5}\frac{GMR^2}{r^3}\Omega.\label{gLf}
\end{equation} 
That is to say, the gravitomagnetic field of massive rotating NS generates a force similar to the Lorentz force in electromagnetic field, which causes a split of the orbital motion of accreted particles, like the classical Zeeman effect in magnetic field. Accordingly, the circular orbital motion on the binary plane changes by the frequency expressed as Eq. (\ref{gLf}) in the vertical direction of accretion disk.

\subsection{Closure orbits and the quasi-quantized disk}

We consider the accretion flows in a turbulent Keplerian disk, 
in which the charged particles, subjected to a Lorentz force and gravity described by Eq. (\ref{force}), obey helical trajectories, 
i.e. open circular orbits at each radius. Because of the rotation of NS, the accreted matter is endowed with two frequencies, i.e. angular frequency in the direction of rotation and gravitational Larmor frequency in vertical direction of disk. The former deviates the orbital motion from a circular orbit, and the latter leads to some vertical split. In this case, the geodesic motion, thus the physical orbits of particles are not holonomy invariant. 
However, if the orbital holonomy invariance can be satisfied at some radii, 
the physical orbits for accreted particles will be closed. 
If and only if the gravitational Larmor frequency and angular velocity of accreted plasma satisfy
\begin{equation}
l\omega(r) = n\omega_L.\label{quant-con},
\end{equation}
where $n$ and $l$ are integers with $n,l \geq1$, the vertical split along disk and deformation in the direction of rotation of NS can be harmonious \cite{2013A&A...552A..10S}, i.e. the vertical oscillation and deformed orbital circular motion is syntonic. As a result, the orbital motion of accreted matter on separated and deformed open circular orbit returns a closed circular one. Because of the appearance of some closed orbits, the accretion disk near the magnetosphere will have
a band of quasi-quantized structure in the gravitoelectromagnetic field 
and axisymmetric space-time, due to the rotation of NS. 
The radii of closure circular orbits are give by
\begin{equation}
r^3 = \frac{4n^2}{l^2}\frac{GM}{\Omega^2}.\label{quanti-radius}
\end{equation}
On these closure orbits, the test particles are in a stable state. 
The dissipation process, i.e. collisions, shocks, and viscosity, and so on, 
in the turbulent and viscous disk, will disturb the stable state, 
which are responsible for the spiral-in and ensure the continuous accretion. 

\section{Summary and Discussion}

In accreting NS/LMXBs, the behavior of accretion flows 
in the inner region of the accretion disk has long been a challenging problem, 
due to the effects of the weak gravitational and strong magnetic fields, 
as well as the rotation of NS. 
In this work, we study the dynamics of accreted plasma in the accretion disk of NS/LMXBs.

We, firstly, deal with the effects of Newtonian gravity and magnetic fields 
by means of canonical Hamiltonian mechanics and describe the classical motion of accretion flows
with Hamiltonian equations of motion. Controlled by the gravity and magnetic field of NS, 
the accreted material obeys Eq. (\ref{cor-Ham}) and Eq. (\ref{mom-Ham}), moving on helical trajectory and spiral-in, 
with Keplerian velocity at each radii in the accretion disk. 
However, due to the rotation of NS, the gravitational potential of NS deviates
from the original spherical symmetry, resulting in axisymmetric space-time metric, 
and the rotational mass-current gives rise to a gravitomagnetic field. 
Both effects have influence on the helical spiral-in of accretion flows, 
contributing to a perturbation in Hamiltonian. 
As a result, the charged particles obtain two additional frequencies, 
i.e., the local angular velocity arising from local frame-dragging 
via oblateness and the gravitational Larmor frequency of plasma in the gravitomagnetic field. The local angular velocity make the orbits of particles deform the original circular orbits, while gravitational Larmor frequency leads to orbital split in vertical direction of disk.
When certain relation described  by Eq. (\ref{quant-con}) between these two frequencies is satisfied, a syntony between the vertical oscillation and the deformed orbital motion occurs, which contributes to some discrete circular closure orbits.
Accordingly, the accretion disk in the inner region shows a band of quasi-quantized structure. 

Moving on the closure circular orbits, the accreted material is in a stable state, 
with first derivative of angular momentum larger than or equal to zero, 
which correspond to a minimum of the effective potential.
With a slight perturbation, the test particle will oscillate around the minimum, 
manifesting as drift of the orbital frequency. 
If the perturbation is strong enough to transfer sufficient angular momentum 
outwardly and drive the particles to leave this state, 
the material will continue to follow the original helical track and spirals in. 
In a turbulent and viscous accretion disk, 
the dissipative process, i.e. viscosity, collisions of elements, shocks, and so on, 
is responsible for the perturbation.

From astrophysical accretion point of view, 
the sudden change of physical environment may give rise to particular phenomenon. 
In the quasi-quantized structure of the inner accretion disk, 
the motion of accretion flows changes from helical infall 
to closure circular orbital motion at some preferred radii, 
which accordingly contribute to special emission. On a closure orbit at radius $r$, 
the plasma moves at Keplerian velocity, radiating X-ray photons 
with an observed flux modulated at the Keplerian frequency $\sqrt{GM/r^3}$. 
However, a slight perturbation due to dissipative processes in the turbulent disk 
will lead to an oscillation around $r$. 
As a result, the original Keplerian frequency at $r$ goes with some drift, 
instead of the exact Keplerian frequency $\sqrt{GM/r^3}$. 
Such phenomenon can manifest as some width of frequency in the Fourier power density spectrum, 
which is referred to as signals of QPOs 
\cite{1989ARA&A..27..517V, 2006csxs.book...39V}. 
In particular, the dynamical timescale of the innermost boundary of an accretion disk 
is the same order of kHz QPOs. In addition, the radial oscillation of plasma, 
with different fixed frequencies on closure orbits, e.g. epicyclic frequencies, 
may result in resonance \cite{2004ApJ...603L..89K, Abramowicz:2004je} 
or certain frequency correlations 
\cite{2006csxs.book...39V, 1999ApJ...520..262P, 2002ApJ...572..392B}. 
In addition, considering that a forced damped harmonic oscillation may appear 
in GEM force and solving the equation of motion for such an oscillator, 
we may obtain photon polarization for different amplitudes 
and different resonant frequencies of the oscillations \cite{Mirza:2003fq}, 
with which we can further investigate some fundamental physics, 
such as CPT symmetry or symmetry violation.

\begin{acknowledgments}

Wang thanks for the anonymous referee's advisements very much. This work was supported by the National Science Council of the Republic of China through grants
NSC 99-2112-M-007 -017 -MY3, 101-2923-M-007 -001 -MY3, and 102-2112-M-007 -023 -MY3.

\end{acknowledgments}

\end{document}